\documentclass[twocolumn]{aastex631}

\received{October 18, 2024}
\accepted{June 28, 2025}

\shorttitle{}
\shortauthors{Wu et al.}
\graphicspath{{./}{figures/}}

\begin{document}

\title{A Common Origin of Normal Type Ia Supernovae Suggested by the Photometric Diversity}

\author{Weiyu Wu}
\affiliation{School of Astronomy and Space Sciences,
University of Science and Technology of China, Hefei, 230026, China}
\affiliation{Department of Astronomy, University of Science and Technology of China, Hefei 230026, China}

\author[0000-0002-9092-0593]{Ji-an Jiang}
\affiliation{Department of Astronomy, University of Science and Technology of China, Hefei 230026, China}
\affiliation{CAS Key Laboratory for Research in Galaxies and Cosmology, Department of Astronomy, University of Science and Technology of China, Hefei, 230026, China}
\affiliation{National Astronomical Observatory of Japan, 2-21-1 Osawa, Mitaka, Tokyo 181-8588, Japan}

\author{Dezheng Meng}
\affiliation{School of Astronomy and Space Sciences,
University of Science and Technology of China, Hefei, 230026, China}
\affiliation{Department of Astronomy, University of Science and Technology of China, Hefei 230026, China}

\author{Zelin Xu}
\affiliation{School of Astronomy and Space Sciences,
University of Science and Technology of China, Hefei, 230026, China}
\affiliation{Department of Astronomy, University of Science and Technology of China, Hefei 230026, China}

\author[0000-0003-2611-7269]{Keiichi Maeda}
\affiliation{Department of Astronomy, Kyoto University, Kitashirakawa-Oiwake-cho, Sakyo-ku, Kyoto 606-8502, Japan}

\author{Mamoru Doi}
\affiliation{National Astronomical Observatory of Japan, 2-21-1 Osawa, Mitaka, Tokyo 181-8588, Japan}
\affiliation{Department of Astronomy, Graduate School of Science, The University of Tokyo, 7-3-1 Hongo, Bunkyo-ku, Tokyo 113-0033, Japan}

\author[0000-0001-9553-0685]{Ken'ichi Nomoto}
\affiliation{Kavli Institute for the Physics and Mathematics of the Universe (WPI), The University of Tokyo, 5-1-5 Kashiwanoha, Kashiwa, Chiba 277-8583, Japan}

\author{Naoki Yasuda}
\affiliation{Kavli Institute for the Physics and Mathematics of the Universe (WPI), The University of Tokyo, 5-1-5 Kashiwanoha, Kashiwa, Chiba 277-8583, Japan}

\author[0000-0001-8253-6850]{Masaomi Tanaka}
\affiliation{Astronomical Institute, Tohoku University, Aoba, Sendai 980-8578, Japan}

\author[0000-0002-4060-5931]{Toshikazu Shigeyama}
\affiliation{Department of Astronomy, Graduate School of Science, The University of Tokyo, 7-3-1 Hongo, Bunkyo-ku, Tokyo 113-0033, Japan}
\affiliation{Research Center for the Early Universe, Graduate School of Science, The University of Tokyo, 7-3-1 Hongo, Bunkyo-ku, Tokyo 113-0033, Japan}

\author[0000-0001-8537-3153]{Nozomu Tominaga}
\affiliation{National Astronomical Observatory of Japan, 2-21-1 Osawa, Mitaka, Tokyo 181-8588, Japan}
\affiliation{SOKENDAI (The Graduate University for Advanced Studies), Mitaka, Tokyo, 181-8588, Japan}

\author[0000-0001-5250-2633]{Željko Ivezić}
\affiliation{Department of Astronomy, University of Washington, Box 351580, Seattle, WA 98195-1580, USA}

\author[0000-0003-2874-6464]{Peter Yoachim}
\affiliation{Department of Astronomy, University of Washington, Box 351580, Seattle, WA 98195-1580, USA}

\author[0000-0001-8738-6011]{Saurabh W. Jha}
\affiliation{Department of Physics and Astronomy, Rutgers, The State University of New Jersey, 136 Frelinghuysen Road, Piscataway, NJ 08854, USA}

\author[0000-0002-1517-6792]{Tinggui Wang}
\affiliation{Department of Astronomy, University of Science and Technology of China, Hefei 230026, China}

\author[0000-0001-7266-930X]{Nao Suzuki}
\affiliation{Department of Physics, Florida State University, Tallahassee, Florida}

\author[0000-0002-6174-8165]{Hisanori Furusawa}
\affiliation{National Astronomical Observatory of Japan, 2-21-1 Osawa, Mitaka, Tokyo 181-8588, Japan}

\author[0000-0001-5576-8189]{Andrew J. Connolly}
\affiliation{Department of Astronomy, University of Washington, Box 351580, Seattle, WA 98195-1580, USA}

\author{Satoshi Miyazaki}
\affiliation{National Astronomical Observatory of Japan, 2-21-1 Osawa, Mitaka, Tokyo 181-8588, Japan}

\correspondingauthor{Ji-an Jiang}
\email{jian.jiang@ustc.edu.cn}




\begin{abstract}

In recent years, with an increasing number of type Ia supernovae (SNe Ia) discovered soon after their explosions, a non-negligible fraction of SNe Ia with early-excess emissions (EExSNe Ia) have been confirmed. In this letter, we present a total of 67 early-phase normal SNe Ia from published papers and ongoing transient survey projects to systematically investigate their photometric behaviors from very early time. We found that EExSNe Ia in our sample have longer rise and brighter peak luminosities compared to those of non-EExSNe Ia. Moreover, EExSNe Ia commonly have ``tiny red-bump" or ``blue-plateau" features in the early $B-V$ color while non-EExSNe Ia show blueward evolution from the very beginning. Here, we propose that the thin-helium double-detonation scenario can phenomenologically explain the photometric diversities of normal SNe Ia considering different white dwarf-He-shell mass combinations and the viewing-angle effect, implying a unified explosion mechanism of normal-type SNe Ia. To further testify the possible common origin of normal SNe Ia, systematical studies of multiband photometric and spectral properties of early-phase SNe Ia through the new generation wide-field time-domain survey facilities and global real-time follow-up networks are highly demanded.

\end{abstract}

\keywords{}


\section{INTRODUCTION} \label{sec:intro}
Type Ia supernovae (SNe Ia), especially for the normal type that consist of $\sim$ 70\% of the whole population \citep{blondin2012spectroscopic} play an important role in cosmological distance measurements \citep{perlmutter1997measurements,perlmutter1999measurements,riess1998observational}. However, the long-standing debate on their origin prevents their accuracy as the cosmic distance indicator. Theoretically, observations within the first few days of the SN explosion could serve as effective probes of their progenitor systems and explosion physics \citep{kasen2009seeing,maeda2018type}. These early observations can provide crucial constraints on the companion type of the progenitor system, the distribution of circumstellar material (CSM) arising from accretion or merger processes, and specific explosion mechanisms \citep{kutsuna2015revealing,jiang2017hybrid,maeda2018type,maeda2023initial}.

Under the single degenerate scenario \citep{2018SSRv..214...67N}, interactions between SN ejecta and a companion star emits radiation at ultraviolet and optical wavelengths in the first few days of the explosion \citep{kasen2009seeing}. Similarly, such a blue early excess (EEx) also can be expected from interactions between SN ejecta and surrounding dense CSM through both single and double degenerate (SD and DD) scenarios \citep{piro2016exploring,jiang2021discovery,maeda2023initial}. The composition of SN ejecta can be different based on specific explosion mechanisms, yielding variable light curve behavior in the early phase. Subsonic mixing \citep{reinecke2002three} and surface helium detonations \citep{jiang2017hybrid,maeda2018type,polin2019observational,2020ApJ...888...80L,2021ApJ...909..152L} through e.g., the double detonation scenario \citep{shen2018sub,tanikawa2018three,tanikawa2019double} can lead to overabundance of radioactive iron-peak density elements. These elements that include \(^{56}\text{Ni}\), \(^{52}\text{Fe}\), and \(^{48}\text{Cr}\) can be found in the outermost layers of SN ejecta, leading to red color evolution in the very early time. Moreover, a growing evidence suggests that a large fraction of EExSNe Ia may originate from the abundant radioactive $^{56}$Ni decay at the surface of SN ejecta through specific explosion mechanisms \citep{jiang2018surface,Miller_2018}.

Thanks to recent high-cadence time-domain surveys with wide-field survey facilities, statistical studies of well-sampled early light curves also indicate that a considerable amount of SNe Ia have EEx features \citep{magee2020determining,deckers2023photometric,fausnaugh2023yearstypeiasupernovae}. \cite{yao2019ztf} constructed a sample of 127 SNe Ia discovered by the Zwicky Transient Facility (ZTF) in 2018. At \( z \sim 0.1 \), the majority of the supernovae exhibit positive light curve shape parameters ($x_1$), indicating a sample bias towards over-luminous, slowly declining supernovae at higher redshifts. \cite{deckers2022constraining} identified 4 normal SNe Ia with EEx features and found that the intrinsic rate of EEx in SNe Ia at z $<$ 0.07 is 18±11\%. \cite{burke2022early} presents a sample of 9 nearby early-phase SNe Ia (z $<$ 0.01), finding that these objects are predominantly near-UV blue. \cite{jiang2018surface} investigated 23 early-phase SNe Ia and found that the rare detection of EEx in normal SNe Ia can be attributed to the intrinsically inconspicuous nature of the EEx emission or a narrower range of observable angles under specific EEx scenarios. 

In contrast to EEx features discovered in different SN Ia subclasses, reliable EEx detections of normal SNe Ia were reported very recently. With the largest normal early-phase SNe Ia up to date, here, we present new evidence that normal SNe Ia may have a common origin based on the photometric diversities from early time. The structure of this letter is as follows: In Section~\ref{sec:data sample}, information of data sources and reduction methods are provided. Statistical results of the light-curve behavior, color, and rise time of the new early-phase SNe Ia sample are given in Section~\ref{sec:results}. Discussions on the photometric diversities and a possible common origin of normal SNe Ia are in Section~\ref{sec:discussion}, and our conclusions are summarized in Section~\ref{sec:conclusion}. 

\section{DATA SAMPLE} \label{sec:data sample}
\begin{figure*}
    \centering
    \includegraphics[width=0.9\textwidth]{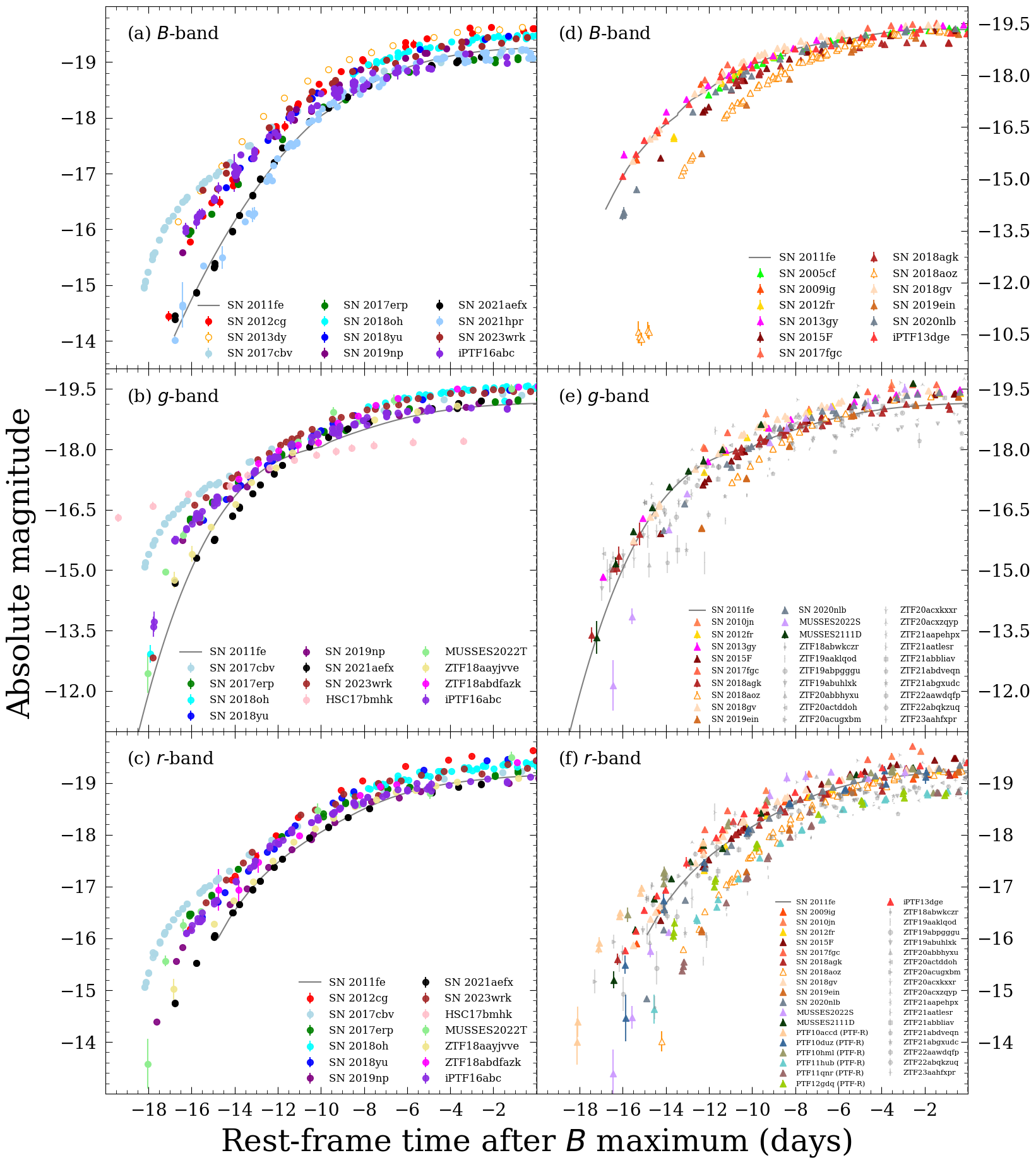}
    \caption{Rising light curves of 54 early-phase normal SNe Ia in either of $B$, $g$, and $r$ bands. EExSN Ia and non-EExSN Ia samples are shown in left and right columns respectively. SN 2013dy and SN 2018aoz are denoted by open symbols. Note that comparisons with the other 13 SNe Ia that only have early-phase light curves in broader bands (e.g., clear and LSQ$gr$, \cite{firth2015rising}) are not applied due to the likely dilution of potential EEx features under the broad-band photometry.}
    \label{fig:image1}
\end{figure*}

We selected 67 normal SNe Ia that have relatively high-cadence photometry (i.e., multiple observations in continuous $\sim3$ days) in the early rising phase from published papers and ongoing survey projects. Here, spectroscopically-classified normal SNe Ia with brightness fainter than $\sim -15.5$ mag upon their first detection in any bands are defined as ``early-phase normal SNe Ia". This sample enables a systematic study of properties of normal SNe Ia in their early phase, as shown in Figure~\ref{fig:image1}.

Firstly, 40 early-phase normal SNe Ia from published papers are summarized, as shown in Table~\ref{tab:events}. Generally, their spectral features and luminosities are in line with those of normal SNe Ia. Two SNe Ia, SN 2012cg and SN 2017cbv, with relatively higher luminosities are included in the comparison of early evolution among normal SNe Ia given the similar spectral features compared with normal SNe Ia. To further increase the sample size, we select 24 early-phase SNe Ia at $z < 0.1$ from spectroscopically normal SNe Ia discovered by ZTF and ATLAS since 2018. In addition, MUSSES 2021 and 2022 campaigns discovered dozens of fast-rising transients (Jiang, J.-a., et al. 2025, in prep), and three early-phase normal SNe Ia, MUSSES2111D, MUSSES2022S, and MUSSES2022T are included to our sample.

\subsection{ZTF and ATLAS Sample} \label{subsec:ZTF and ATLAS sample}
ZTF allocates 40\% of its time to conduct a public survey of the Northern Sky. Public data are acquired using $g$ and $r$ filter pairs, with a preferred cadence of once every three/two nights. In this study, high-quality $g_{ZTF}$ and $r_{ZTF}$ (hereafter referred to as $g$ and $r$) light curves of SNe Ia from 2018 to 2023 are used for detailed analysis.

The forced PSF photometry method developed by \cite{yao2019ztf, masci2023newforcedphotometryservice} was utilized to acquire the initial sample of ZTF normal SNe Ia, totally 5372 events. Following the protocol suggested by \cite{yao2019ztf}, observations with high reduced chi-square statistics ($\chi^2_{\nu} > 4$) or significant baseline offsets ($|C| > 15$) were excluded. For ZTF SNe, normal SNe Ia discovered at least 10 days before peak with absolute magnitude of $\gtrsim -15.5$ mag are classified as early-phase normal SNe Ia, which contribute a total number of 22 to our sample.

Galactic extinction corrections were applied using the dust maps by \cite{schlegel1998maps}, assuming $R_V = 3.1$. $K$-corrections for SNe Ia are uncertain due to extrapolation from 15 days before maximum brightness to earlier times. Moreover, \cite{bulla2020ztf} observed that $K$-corrections for a subset of the ZTF SNe Ia sample are minor, with $K$-corrections for the objects with the highest redshift being $\leq 0.1$ mag. Thus, $K$-corrections are not applied in the subsequent analysis.

The ATLAS Forced Photometry service was applied for getting complete photometric information of total $\sim$ 1300 SNe discovered by ATLAS \citep{shingles2021release} from 2016 to 2023. From the ATLAS SN sample, 54 early-phase SNe Ia were selected with $c$ or $o$-band magnitude of $\gtrsim -15.5$ mag at first detection. All SNe are spectroscopically classified as normal SNe Ia using the SuperNova Identification tool (SNID; \cite{blondin2007determining}). After excluding SNe without multiple observations in continuous $\sim3$ days of their rising phases, 7 early-phase normal SNe Ia were finally selected.

\subsection{MUSSES Sample} \label{subsec:MUSSES sample}
Our sample is further enlarged through the Multi-band Subaru Survey for Early-phase SNe Ia (MUSSES; \cite{jiang2017hybrid,jiang2022musses2020j} project. MUSSES has one or two observing campaigns every year, and each campaign include two parts: a Subaru/HSC survey (several half-nights) and follow-up observations. In the HSC survey, Subaru/HSC observed dozens of square degree sky nightly, with a typical $g$-band limiting magnitude of about 26.0 (5$\sigma$), to search for early-phase SNe Ia with complete multi-band photometric information.

Astronomical transients were identified using the HSC transient pipeline and machine learning techniques \citep{jiang2017hybrid,yasuda2019hyper,Jiang_2020}. Initially, objects with ``negative'' detections, such as fading transients, or those with point-source hosts at the same location, like variable stars, were filtered out. Remaining candidates were cross-checked by team members. Based on specific criteria, such as proximity to the host, non-association with X-ray sources, and light curve shapes, about one thousand transients were classified as SN candidates and three of them are finally identified as early-phase SNe Ia through MUSSES follow-up network from MUSSES campaigns in Nov 2021 and Oct 2022. Multi-band imaging follow-ups were conducted with APO 3.5m telescope spanning approximately $-8$ to $+40$ days after the $B$-band maximum. Spectroscopic follow-ups were triggered by the 9.2-m Southern African Large Telescope (SALT) and the 8.1-m Gemini-North telescope at specific epochs to capture the spectral evolution from approximately $-2$ days to 1 month after the $B$-band maximum (Jiang et al. 2025, in prep). Details of the three early-phase normal SNe Ia, MUSSES2111D, MUSSES2022S, and MUSSES2022T are summarized in Table~\ref{tab:events}.

For the HSC data, standard image reduction procedures, including bias, dark, flat-field, and stripe correction, as well as astrometric measurements and photometric calibration against the Pan-STARRS1 (PS1) 3$\pi$ catalog \citep{Tonry_2012,Magnier_2013}, were performed using the HSC pipeline, a version of the Large Synoptic Survey Telescope (LSST) stack \citep{axelrod2010open,bosch2018hyper,ivezic2019lsst}. Image subtraction was applied using deep co-added reference images created from data taken in the past. Photometry was performed on the positions of objects left in the reference-subtracted images using forced point spread function (PSF) photometry. For more detailed information on the HSC data reduction, refer to \cite{aihara2018hyper} and \cite{yasuda2019hyper}.

The APO 3.5m multi-band images were reduced in a standard manner for the CCD photometry. For the z and i bands of ARC data, fringes were subtracted. Position calibration was performed using SCAMP \citep{2006ASPC..351..112B}. Pan-STARRS1 (PS1) 3$\pi$ catalog was used as the reference catalog to exclude obvious variable stars in the flux field and to derive zero-point fluxes by comparison with reference stars. After stacking images from the same day, Pan-STARRS was used as the reference image for subtraction using Saccadic Fast Fourier Transform (SFFT, \cite{hu2022image}), and aperture photometry was performed on the differential images using SExtractor \citep{1996A&AS..117..393B}.

\section{RESULTS} \label{sec:results}
\subsection{EExSNe Ia}
Following the definition from \cite{jiang2018surface}, an early-phase normal SNe Ia with additional excess emissions compared to a smooth rise is defined as an EExSN Ia in this paper. Normal SNe Ia, whose rising light curves can be fitted by a power-law function when reach half of the normalized peak flux, are defined as non-EExSNe Ia. EEx emissions in our sample are systematically investigated based on early light curve behavior. Specifically, we examine EEx flux in each sample by applying a power-law light-curve fit (\( F = k (t-{t_0})^{\alpha} \)) from the discovery to $\sim10$ days before the peak. For normal SNe Ia with excess emissions, a distinct bump is clearly observed in the light curve, and the fitting residuals manifest a quasi-sinusoidal pattern. Following the method from \citep{Deckers_2022}, residuals of two consecutive detections in the same band are both non-zero, with at least one residual exceeding 2\% of the peak flux, an additional Gaussian component is applied to fit extra emissions in the early rising phase of EExSNe Ia.

As for the 40 published early-phase SNe Ia, we suggest reclassifying the early light curve behaviors of SN 2013dy and SN 2018aoz for the following reasons. SN 2013dy was firstly discovered by Lick Observatory Supernova Search \citep{stahl2019lick}. Due to the difficulty of early light curve fitting with a classical power-law model, \cite{zheng2013very} applied a broken power law function to fit the light curve in the first 8 days. As can be seen in Figure~\ref{fig:image1}, compared to notable EExSNe Ia such as SN 2012cg, SN 2017erp, and SN 2018yu, the light curve of SN 2013dy (represented by orange open circles) indeed has a more prominent EEx feature, which has been supported by recent work \citep{Zhai_2016}. We thus reclassify SN 2013dy as an EExSN Ia.

In the very early phase of SN 2018aoz, a spike deviating from a simple power-law fit can be marginally distinguished in the $V$ and $i$ bands \citep{ni2022infant}. In the subsequent rapid rise phase, the power law fit appropriately matches the rising light curves, consistent with the majority of normal SNe Ia such as SN 2011fe (Figure~\ref{fig:image1}). Given the inconspicuous excess emission that occurred only in the very early phase, where no other SNe Ia has been observed, and the generally smooth rising behavior of SN 2018aoz (orange open triangles in Figure~\ref{fig:image1}), we thus classify it as a ``non-EEx" normal SN Ia.

As for the additional ZTF and ATLAS samples (24 in total), we do not see early excess considering the uncertainties associated with individual flux measurements. However, as shown in Figure~\ref{fig:image1}, a clear early-bump feature can be identified in MUSSES2022T among three MUSSES samples. 

In total, 15 SNe show EEx features among 67 early-phase normal SNe Ia. Rest-frame rising light curves relative to the $B$-band peak in $B$, $g$, and $r$ bands are given in Figure~\ref{fig:image1}. The ratio between EExSNe Ia and non-EExSNe Ia of 17 SNe Ia at $z<0.01$, of which the most comprehensive early data were obtained, is almost the same. However, it is unclear whether such a high EEx ratio is related to a specific EEx scenario or due to potential selection bias for low-$z$ SNe from literature (e.g., SNe with peculiar light-curve behavior are more likely to be published). A large unbiased sample from future high-cadence transient surveys are highly needed to stringently constrain the EEx ratio of normal SNe Ia.

\subsection{Rise Time}
Given that the rise time before SNe Ia reach a brightness of -13 mag is negligible and will not be strongly influenced by any EEx mechanisms, we use $t_{m(-13)}$ as the zero point for calculating the rise time (i.e., $\Delta{t_{m(-13)}}$) of our early SNe sample. A single power-law fit was applied to determine $\Delta{t_{m(-13)}}$ in the $B$, $g$, or $r$ band of non-EExSNe Ia\footnote{Compared to the definition of rise time in previous studies, which is the time from the extrapolated explosion time to the peak, the rise time $t_{m(-13)}$ defined here is shorter than this duration. In addition, as can be seen in Figure~\ref{fig:rise_gr_appendix}, the derived $\Delta{t_{m(-13)}}$ in $g$ and $r$ bands usually show a good agreement except for a few cases where the rise times can be quite different, suggesting a low wavelength sensitivity of the parameter.}. The prior parameter range for the power-law index we used is $1 < \alpha < 3$. However, a single power-law cannot fit $\Delta{t_{m(-13)}}$ if SNe Ia have EEx emission, especially for the bump-like excesses \citep{jiang2018surface}. Therefore, for EExSNe Ia, an additional Gaussian component was used to fit EEx feature. The $lmfit$ package is used to perform a least squares fit to obtain initial fitting parameters, which are then refined using Markov chain Monte Carlo (MCMC) based on the derived parameters \citep{newville2023lmfit}. For MCMC sampling, uniform prior distributions were adopted for the parameters, assuming equal probability within the defined ranges without prior bias toward specific values. 

\begin{figure*}
    \centering
    \includegraphics[width=0.75\textwidth]{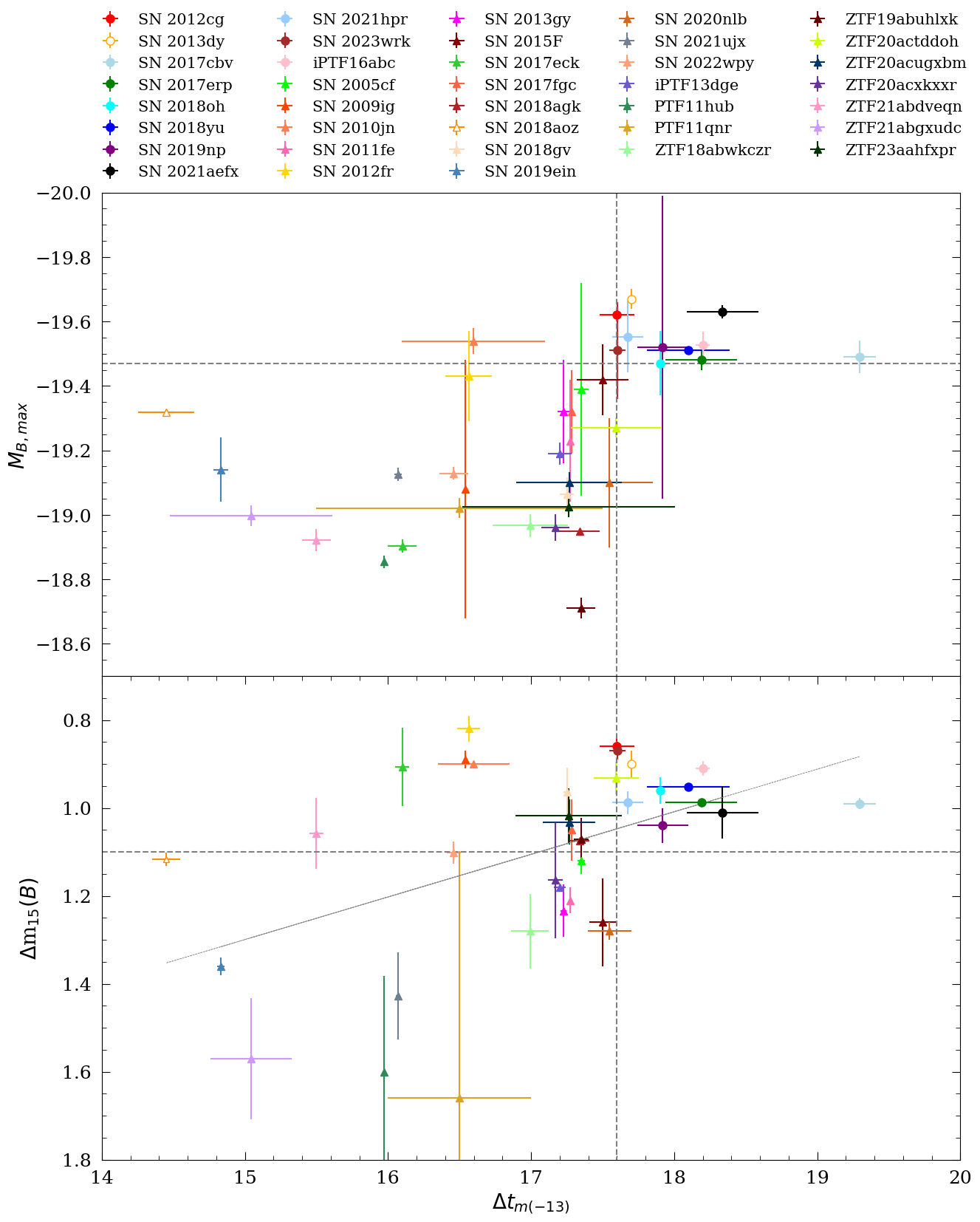}
    \caption{The $B$-band peak absolute magnitude ($M_{B,\text{max}}$) versus $\Delta m_{15}(B)$ and $\Delta{t_{m(-13)}}$ of normal SNe Ia at {$z<0.03$}. EExSNe Ia are denoted by circles. $M_{B,\text{max}}$ of 3 ATLAS (SN 2017eck, SN 2021ujx, SN 2022wpy) and 8 ZTF discovered non-EExSNe Ia are derived from SALT2 fitting. Grey dashed lines correspond to specific values of $\Delta{t_{m(-13)}}$, $M_{B,\text{max}}$ and $\Delta m_{15}(B)$ that separate our EExSN Ia and non-EExSN Ia samples into different quadrants. It is clear to see that all EExSNe Ia are confined to the upper right corner. A linear fit is shown as a grey dotted line in the bottom panel. SN 2013dy and SN 2018aoz are denoted by open symbols.}
    \label{fig:rise}
\end{figure*}

\begin{figure*}
    \figurenum{A1}
    \centering
    \includegraphics[width=0.95\textwidth]{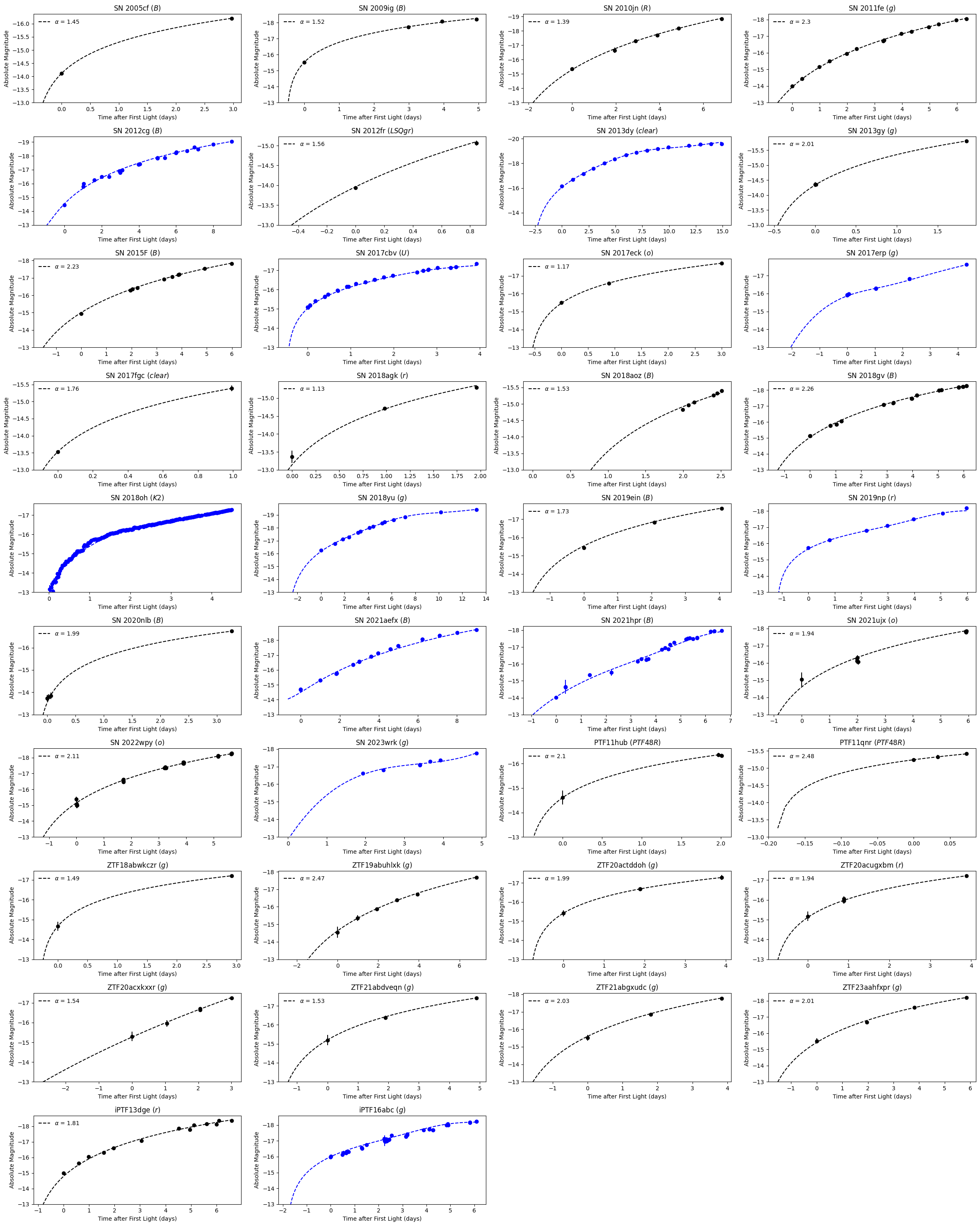}
    \caption{Rising light-curve fittings of 38 early-phase normal SNe Ia at $z<0.03$. Light curves of 11 EExSNe Ia are shown in blue. Power-law fitting indices of non-EExSNe Ia are shown in each sub-panels.}
    \label{fig:rise_appendix}
\end{figure*}

\begin{figure*}
    \figurenum{A2}
    \centering
    \includegraphics[width=0.95\textwidth]{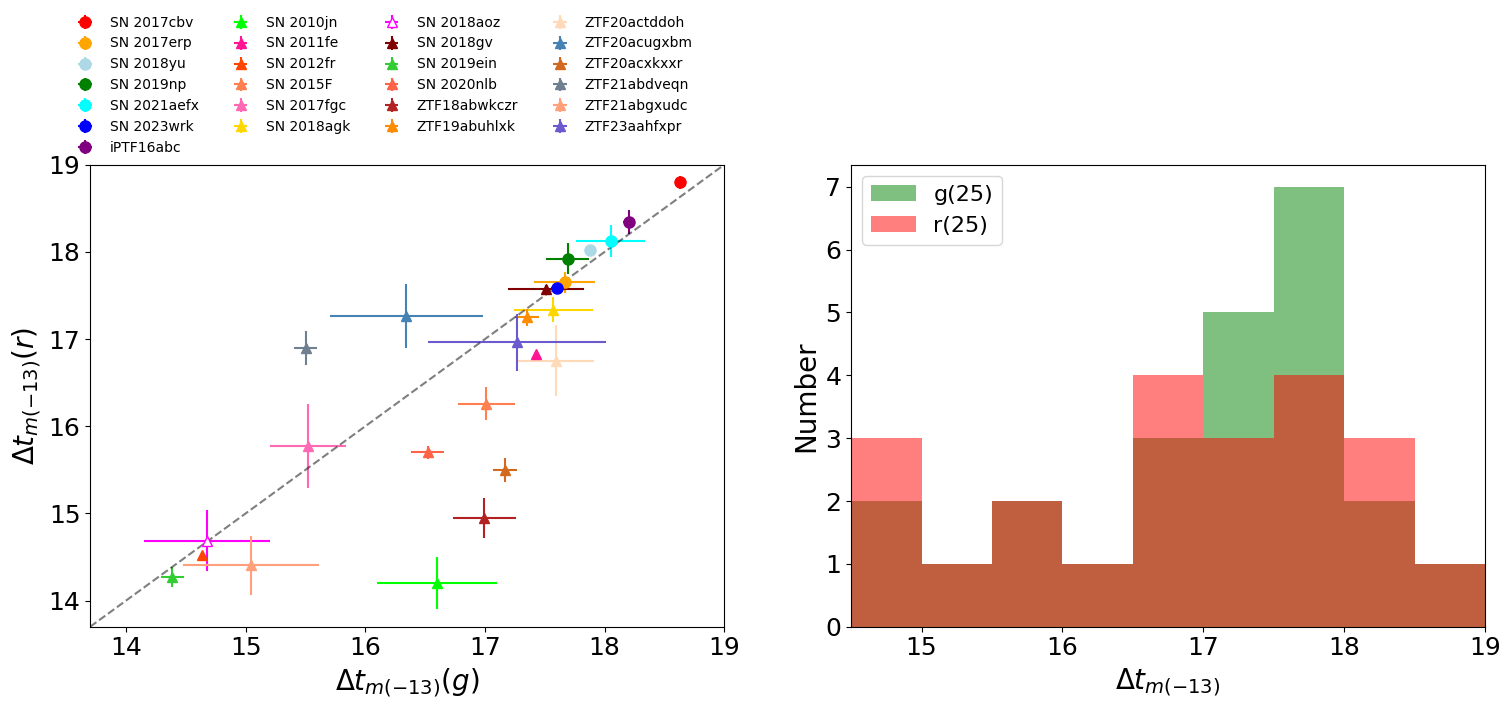}
    \caption{A comparison of $g$- and $r$-band $\Delta{t_{m(-13)}}$ of normal SNe Ia (left panel) and their number distributions (right panel). The derived $\Delta{t_{m(-13)}}$ in two bands show quite good consistency for most early-phase SN Ia sample given in Figure~\ref{fig:rise}.}
    \label{fig:rise_gr_appendix}
\end{figure*}

As redshift increases, EEx features are hardly to be identified due to the shallow depth of optical surveys led by $\lesssim$ 1m class telescopes (a typical EEx peak range is about 19 – 20 mag at $z \sim 0.03$). Therefore, a total of 38 SN samples at $z<0.03$ are included in order to ensure a good photometric dynamical range and sample size for the rise time statistics. The top panel of Figure~\ref{fig:rise} shows the correlation between the rise time and the peak absolute magnitude for our sample. This result indicates that EExSNe Ia generally have longer rise time and higher peak brightness than those of non-EExSNe Ia. Notably, the light curve of SN 2013dy is well-fitted by incorporating a Gaussian component with a power law, yielding the rise time as long as $\sim17.7 \pm 0.05$ days. Our fitting result further indicates the presence of excess emission in the early time of SN 2013dy. In addition, the rise time of SN 2018aoz derived from a power-law fitting is only $14.45 \pm 0.195$ days, indicating a very rapid rise with a untypical subtle ``EEx emission". As can be seen in Figure~\ref{fig:rise}, EExSNe Ia have $\Delta m_{15}(B)$ range of $<1.1$, with $M_{B,\text{max}} > -19.48$, densely assembled at the upper right corner of the graph. $\Delta m_{15}(B)$ of non-EExSNe Ia is more dispersed, while their $\Delta{t_{m(-13)}} < 17.6$ are well categorized on the left side of 17.6 day. Fitting results of 38 SNe at $z<0.03$ are shown in Figure~\ref{fig:rise_appendix}.

\subsection{Color Evolution} \label{subsec:tables}
\cite{bulla2020ztf} claimed that there is no clear difference between red and blue samples categorized by early $B-V$ colors in the early $g-r$ colors \citep{stritzinger2018red}, suggesting that the early color evolution of SNe Ia might be relatively homogeneous in $g$ and $r$ filters. Therefore, the largest contrast between the two classes is captured by $B-V$ colors, while $g-r$ colors tend to wash out the observed spectral differences. It suggests that $B$ and $V$ filters might be the better choice to test different models affecting the early-time colors. In our sample, 22 SNe Ia with both $B$- and $V$-band data were used for further investigation of early color behavior.

\begin{figure*}
    \centering
    \includegraphics[width=0.75\textwidth]{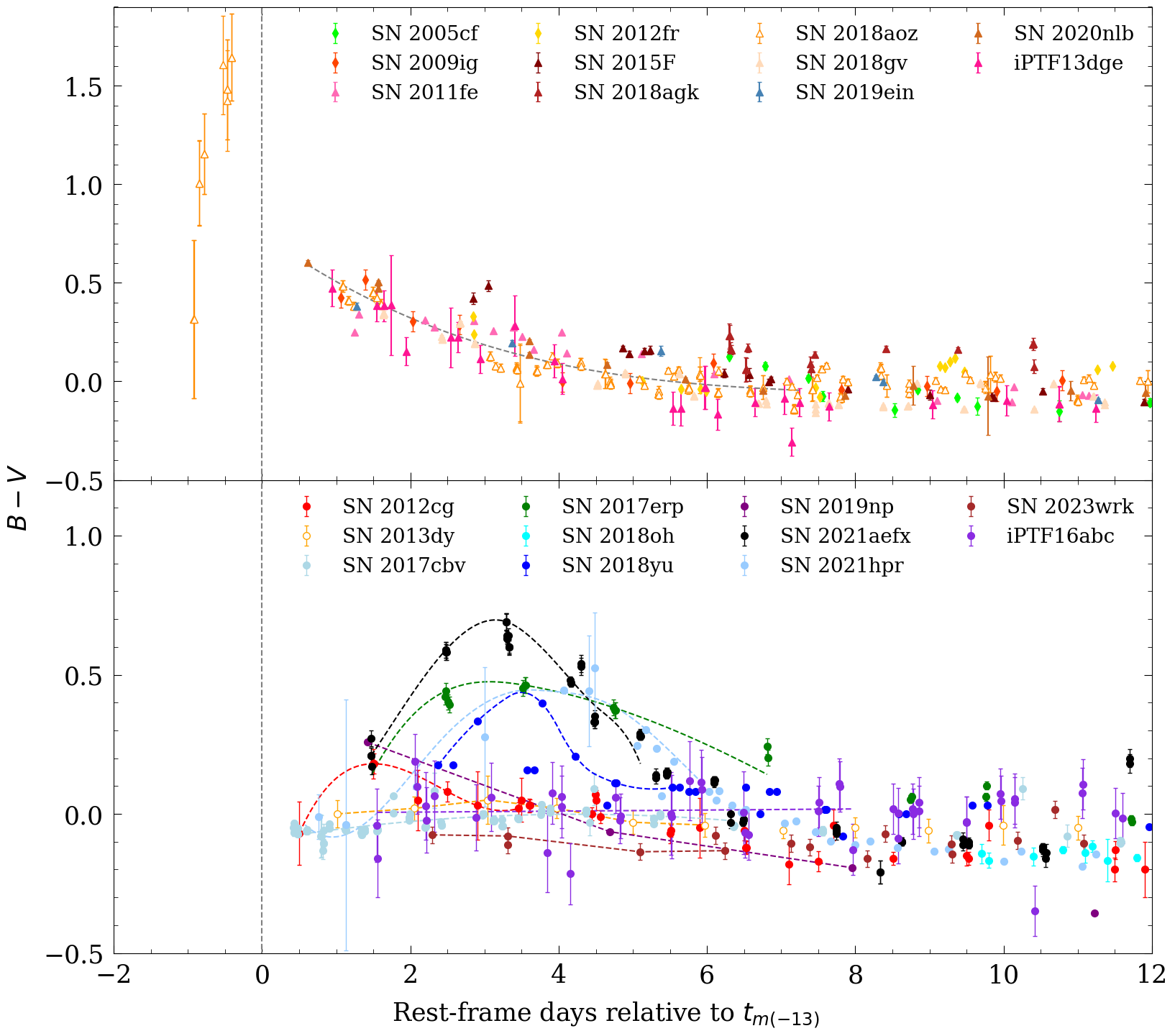}
    \caption{Early $B-V$ color evolution of 22 normal SNe Ia. Symbols are same with those used in Figure~\ref{fig:rise}. The upper panel shows early $B-V$ color evolution of 11 non-EExSNe Ia. A dashed gray curve represents an overall trend of their early color behavior. Early $B-V$ color evolution of 11 EExSNe Ia fitted by dashed curves are given in the lower panel.}
    \label{fig:color}
\end{figure*}

The $B-V$ color evolution relative to $t_{m(-13)}$ of 22 early SNe Ia is shown in Figure~\ref{fig:color}. A clear dispersion is observed in the early phase. Starting from about 6 days after $t_{m(-13)}$, the distribution of $B-V$ values becomes quite uniform, consistent with the results of \cite{stritzinger2018red}. The early $B-V$ color is bluer for the brighter samples ($< -19.5$ mag). In particular, EExSNe Ia generously show blue--red--blue or continuous blue evolution in early phase (i.e., ``tiny red bump" or ``blue plateau"), which is consistent with predictions from \cite{shen2021multidimensional}. Unlike EExSNe Ia, a monotonically blueward early color evolution is found for non-EExSNe Ia.

\section{DISCUSSION} \label{sec:discussion}

\subsection{Photometric diversities in normal SNe Ia}
Previous studies have shown that the average rise time of normal SNe Ia is approximately 17-18 days \citep{firth2015rising,wang2024flight}, though recent ZTF studies found a broader rise-time distribution with an average rise time of $\sim$18.5 days \citep{miller2020ztf}. Systematically investigating the correlation between rise time and blue EEx features of SNe Ia is crucial to find out the origin of ambiguous early excess, as demonstrated for HSC17bmhk by \cite{Jiang_2020}. In recent years, an increasing number of normal SNe Ia with long rise time have been discovered, suggesting that outermost ejecta layers of a considerable numbers of normal SNe Ia may be enriched with $^{56}$Ni. As a result, a higher fraction of EExSNe Ia can be expected, which indeed is in line with the rise-time statistics for our normal SNe Ia sample. 

\cite{stritzinger2018red} reported a red group with a rapid transition from red to blue, and a blue group with bluer and slower evolution, based on the $B-V$ early color evolution of 13 SNe Ia. They argued that SNe in the blue group are predominantly over-luminous and associated with the Shallow Silicon (SS) spectral type, while those in the red category more commonly align with the Core Normal (CN) or CooL (CL) types as classified by \cite{branch2006comparative}. Recently, \cite{han2020sn} included five normal SNe Ia based on \cite{stritzinger2018red} and also found the distinction between red and blue groups in their sample. In terms of the proposed red and blue color groups, our \(B-V\) color statistics of 22 early normal SNe Ia shows distinct color evolution between EEx and non-EExSNe Ia. In particular, ``tiny red-bump" or ``blue-plateau" color behavior for all EExSN Ia sample indicates that photometric diversities of early-phase normal SNe Ia are likely related to the EEx behavior and might be interpreted by a specific EEx mechanism.

\subsection{Thin He-shell double detonation, a common origin for normal SNe Ia?}

The He-shell double detonation and the surface $^{56}$Ni-decay are two popular EEx scenarios that lead to specific elements at the outermost region of SN ejecta \citep{jiang2017hybrid,piro2016exploring,magee2020determining}. In the context of the surface $^{56}$Ni-decay scenario, blue EEx emission can be expected and the strength of which decreases toward the smaller core mass. These mechanisms are indeed not mutually exclusive, and the surface $^{56}$Ni-decay (as a product of the core nucleosynthesis) may dominate over the contribution from the He-ash in the context of the double-detonation model -- this is the scenario proposed here. Moreover, asymmetric $^{56}$Ni distribution can lead to a bumpy EEx instead of a smooth one under the surface $^{56}$Ni-decay scenario \citep{magee2020determining}. As for the thin He-shell double detonation scenario (DDet), adsorptions from intermediate mass elements such as Ti and Ca yield red EEx emission for normal-brightness SNe Ia \citep{jiang2017hybrid,maeda2018type,polin2019observational}. A general trend is expected where the detonation is initiated at a smaller He shell mass for a more massive WD core, for given shell detonation criteria; the deeper gravitational potential for a more massive core will lead to either higher density or temperature, or both, for a given shell mass. While the He shell mass at the initiation of the detonation should depend on the mass transfer history, this trend is expected both in the accreting WD (e.g., \citealp{2007ApJ...662L..95B}) and the WD-WD merger \citep{2022ApJ...941...87I} cases. Therefore, we expect that increasing the WD core mass results in the smaller amount of the He detonation products, and finally EEx powered by the surface $^{56}$Ni-decay scenario will overcome the He-det-induced EEx. In the following, a thin He-shell mass of \(\lesssim 0.02 M_{\odot}\) is assumed to explain a reasonable M$_{B,max}$ for normal SNe Ia\footnote{Note that the accreting white dwarf (WD) scenario generally results in even more massive He shell, except for very massive WDs (\(\gtrsim 1.2 M_{\odot}\); \citealp{kromer2010double}). The double-degenerate (DD) scenario may achieve the thin He shell (e.g., \citealp{shen2021multidimensional}), though its feasibility remains uncertain (e.g., \citealp{2022ApJ...941...87I}).}.

According to the multi-dimensional simulations of \cite{Shen_2021}, for a CO WD with a massive core (e.g., $\sim1.1 M_{\odot}$) and a very thin helium shell (0.0084-0.011 $M_{\odot}$), a promising amount of \(^{56}\text{Ni}\) (0.75-0.76 $M_{\odot}$) can be generated by the double detonation. Phenomenologically, blue EEx emission powered by the surface \(^{56}\text{Ni}\)-dacay with brighter/bluer SNe Ia thus can be expected by further increasing the total mass of WD under the thin He-shell DDet scenario. Furthermore, the viewing angle effect likely dominates the observational diversities among normal SNe Ia. In the following, we illustrate that the photometric behavior of our normal SNe Ia sample can be reasonably explained after taking into account the viewing angle effect through the thin He-shell DDet scenario.

\begin{figure*}
    \centering
    \includegraphics[width=0.75\textwidth]{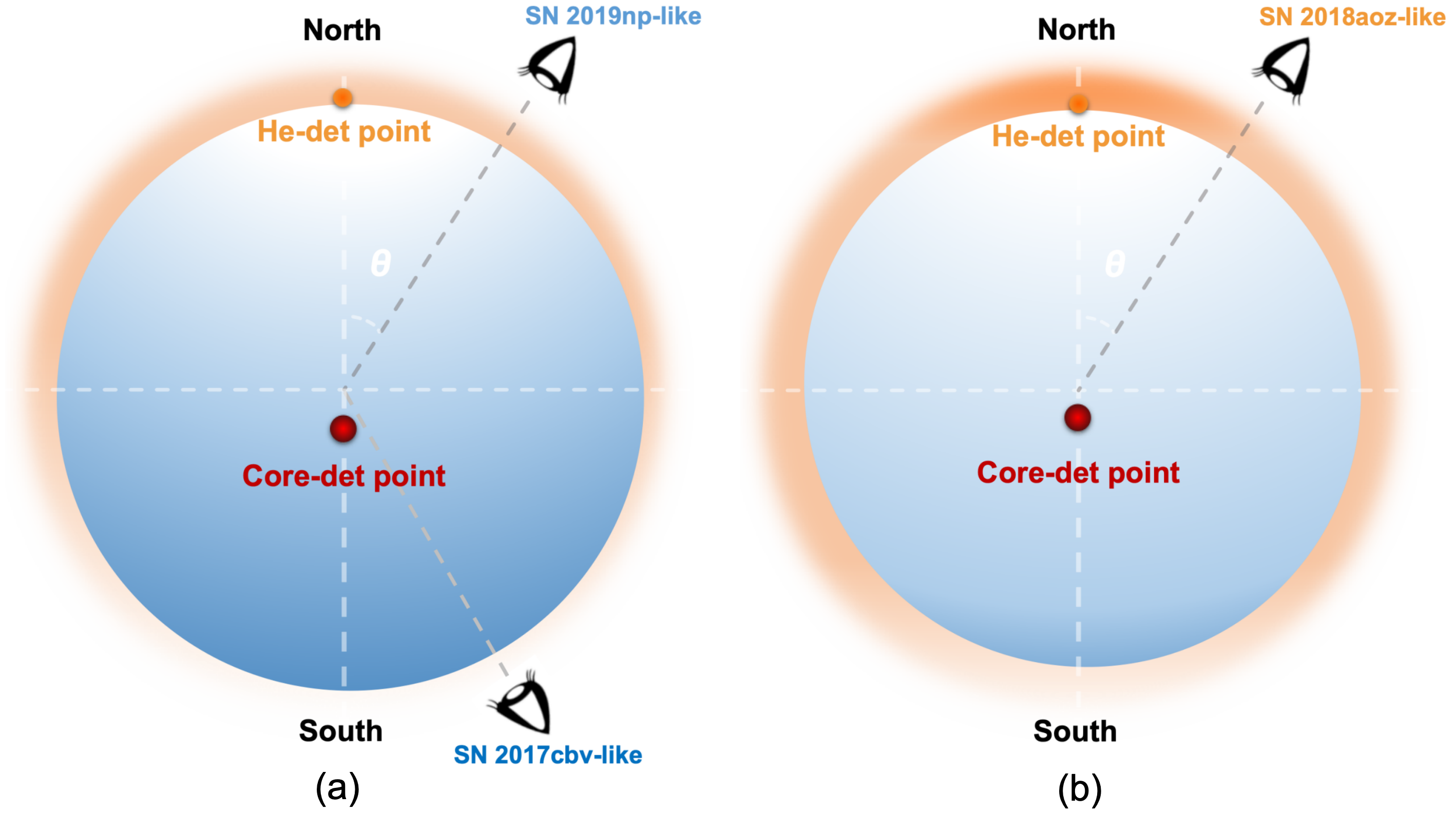}
    \caption{Schematic diagram of the spatial position distribution of early-phase SNe Ia during the explosion within the progenitor system, with key points labeled for orientation. The circle represents the explosion site, with the North and South directions marked. The He shell detonation (He-det) point is shown at the top (North), and the Core-detonation (Core-det) point is on the opposite side to the center (i.e., south). The angle $\theta$ indicates the position angle relative to north. Panels (a) and (b) illustrate SNe with relatively high and low mass, respectively.}
    \label{fig:model}
\end{figure*}

Assuming that the helium shell is ignited at the north pole \citep{2020ApJ...888...80L,2021ApJ...909..152L} (Figure~\ref{fig:model}), the subsequent helium detonation propagates along the WD surface towards the south direction, releasing a shock wave that enters the core and ultimately triggers the carbon core detonation near the symmetric axis of the south pole. The second detonation propagates back and burns through most of the core of a CO WD. Statistically, bolometric luminosities of EExSNe Ia are higher than those of non-EExSNe Ia, which can be naturally interpreted by the difference of \(^{56}\text{Ni}\) amount, and thus related to the mass of WD progenitors. At the meanwhile, the distribution of \(^{56}\)Ni produced by the core detonation can be extended to the outermost layer, yielding a more prominent ``blue" EEx emission instead (i.e., the observed ``EExSNe Ia" with generally blue color around $t_{m(-13)}$, Figure~\ref{fig:color}). As the He-shell mass increases (but still under the thin He-shell DDet scenario), we may expect an inconspicuous EEx emission powered by He-shell detonation that is too fast/faint to be easily captured by ongoing survey projects (i.e., the observed ``non-EExSNe Ia"). To evaluate the influence on light curve diversities caused by the viewing angle effect, two WD core-mass conditions are considered for simplicity (panels (a) and (b) in Figure~\ref{fig:model}, respectively).

Massive WDs with thinner He-shells are schematically illustrated in panel (a) of Figure~\ref{fig:model}. Multidimensional simulations from \cite{boos2021multidimensional} indicate that more abundant \(^{56}\)Ni can be generated near the southern side, thus a prominent blue EEx and longer rise time (e.g., SN~2017cbv-like) would be expected under the surface \(^{56}\)Ni-decay scenario by viewing from the south direction. At the meanwhile, SNe with inconspicuous blue EEx emissions and relatively short rise time would be found near the northern side (e.g., SN~2019np-like). In addition to the above two extreme cases, other EExSNe Ia can be qualitatively interpreted at specific viewing angles. On the other hand, the discovered absolute magnitude of non-EExSNe Ia SN~2018aoz was as faint as $\sim -10$ mag, with an inconspicuous excess appearing in subsequent observations, implying a slightly thicker He-shell\footnote{Note that all discussions here are based on thin He-shell DDet assumption, SN~2018aoz may have a thicker He-shell shell than those of most normal SNe Ia but it is still within the context of the thin-shell scenario.} DDet origin of the SN. In such a case, non-EExSNe Ia can be explained by DDet of less massive WDs under different viewing angles (panel (b) of Figure~\ref{fig:model}). As for the early color evolution of EExSNe Ia, previous studies suggest that Fe-group lines can be suppressed by the high ionization in the earliest phase, leading to the ``red-bump" feature under the He-shell DDet scenario \citep{maeda2018type}; the He-shell detonation products not only provide a source of the line opacities but ionization power through radioactive decay. However, this additional ionization source by the He-detonation products is not sufficient to explain the range of the color behaviors found in the present work; it does not lead to the ``blue plateau", and the predicted ``red bump" under the thin He-shell condition could be much more prominent (e.g., a $B-V$ peak $\sim$ 2) compared to that of the normal EExSNe Ia sample shown here. This could be remedied by introducing an additional source of ionization, i.e., surface \(^{56}\)Ni as proposed here. In addition, the viewing-angle effect could result in the range of the early color behaviors, including the blue-plateau and tiny red-bump color features (i.e., a $B-V$ peak $\lesssim$ 0.7 within a week of the first light).

In Figure~\ref{fig:rise}, there are a few non-EExSNe Ia with smaller $\Delta m_{15}(B)$ and fainter M$_{B,max}$. However, given their M$_{B,max}$ have large uncertainties from literature and the $\Delta m_{15}(B)$ variance range for less-luminous normal SNe Ia indeed is larger than that of luminous normal SNe Ia in the Phillips relation\citep{hicken09}, the physical origin of the possible large $\Delta m_{15}(B)$ variance for non-EExSNe Ia is beyond the scope of this paper and we leave it as a future work. There is a hint that EExSNe Ia tend to be brighter than non-EExSNe Ia for given $\Delta m_{15}(B)$. While this could be contaminated errors in the estimate of M$_{B,max}$ and $\Delta m_{15}(B)$, it may also link to the origin of the EEx feature. For example, it might not be surprising that the peak magnitude may be viewing angle-dependent if the ejecta are asymmetric \citep{maeda2011}, as involved in our proposed scenario explained above. Photometrically, 91T-like SNe Ia are characterized by bright, slowly evolving light curves, similar to EExSNe Ia in our sample. Intuitively, the detonation of an ultra-thin helium shell might explain early excess observed in 91T-like SNe Ia, thereby implying a possible common origin for both 91T-like and normal SNe Ia. However, whether the ultra-thin helium-shell detonation can be realized under the DDet scenario and the necessity of introducing such an extreme condition are not clear \citep{2022ApJ...941...87I,2024PhRvL.133l1201I}. Alternatively, an efficient detonation occurring at the surface of a WD through the gravitationally confined detonation scenario can also yield sufficient \(^{56}\text{Ni}\) to power blue EEx emissions and inconspicuous iron group element absorptions in early spectra of 91T-like SNe Ia \citep{jiang2018surface,jordan2012detonation}. Therefore, further observational evidence is needed to figure out the intrinsic connection between normal and 91T-like SNe Ia.

\section{CONCLUSIONS} \label{sec:conclusion}
In this paper, we build the largest early normal SNe Ia sample from published papers and different survey project. Among the total 67 SNe Ia, all 15 EExSNe Ia have longer rise time and brighter peak absolute magnitude compared to those of non-EExSNe Ia, and the fraction of EExSNe Ia and non-EExSNe Ia without considering a potential selection bias is comparable for the local sample at $z < 0.01$. Besides, EExSNe Ia commonly show ``tiny red-bump" or ``blue-plateau" color feature in the early time while non-EExSNe Ia show blueward evolution from the very beginning. In general, we find that the thin-helium DDet scenario can reasonably explain photometric diversities of normal SNe Ia by altering the core mass of WD and viewing angle.

It is worth noting that though the thin-helium DDet scenario can qualitatively explain the photometric behaviors of normal SNe Ia from early time, detailed analyses especially for the possible connection between the viewing angle effect and spectral diversities of normal SNe Ia should bring key evidence on the DDet scenario proposed here. From March 2024, the 2.5-m Wide Field Survey Telescope (WFST or ``Mozi") jointly constructed by the University of Science and Technology of China (USTC) and the Purple Mountain Observatory (PMO) carried out a 4-month pilot survey (Jiang, J.-a., et al. 2025, in prep) and a remarkable 6-year time-domain survey project started in
 December 2024 \citep{Wang_2023}. In the coming years, SNe Ia discovered by WFST and Vera. C. Rubin/LSST \citep{ivezic2019lsst} will significantly enlarge the sample of well-observed early normal SNe Ia. Systematical studies on early multiband photometric and spectral properties of WFST/LSST SNe would answer the long-standing diversity issue in normal SNe Ia.
\\

\noindent\textbf{Acknowledgements}

We thank the anonymous referee for helpful and constructive comments and suggestions. This work was supported by the National Natural Science Foundation of China (Grant No. 12393811), National Key R\&D Program of China (Grant No. 2023YFA1608100), and the Strategic Priority Research Program of the Chinese Academy of Science (Grant No. XDB0550300). J.J. acknowledges the supports from the Japan Society for the Promotion of Science (JSPS) KAKENHI grants JP22K14069. K.M. acknowledges support from the Japan Society for the Promotion of Science (JSPS) KAKENHI grant JP24KK0070 and 24H01810. K.N. has been supported by the World Premier International Research Center Initiative (WPI), MEXT, and KAKENHI grants JP20K04024, JP21H04499, JP23K03452, and JP25K01046.

The Hyper Suprime-Cam (HSC) collaboration includes the astronomical communities of Japan and Taiwan, and Princeton University. The HSC instrumentation and software were developed by the National Astronomical Observatory of Japan (NAOJ), the Kavli Institute for the Physics and Mathematics of the Universe (Kavli IPMU), the University of Tokyo, the High Energy Accelerator Research Organization (KEK), the Academia Sinica Institute for Astronomy and Astrophysics in Taiwan (ASIAA), and Princeton University. Funding was contributed by the FIRST program from Japanese Cabinet Office, the Ministry of Education, Culture, Sports, Science and Technology (MEXT), the Japan Society for the Promotion of Science (JSPS), Japan Science and Technology Agency (JST), the Toray Science Foundation, NAOJ, Kavli IPMU, KEK, ASIAA, and Princeton University. 
This paper makes use of software developed for the Large Synoptic Survey Telescope. We thank the LSST Project for making their code available as free software at http://dm.lsst.org

The Pan-STARRS1 Surveys (PS1) have been made possible through contributions of the Institute for Astronomy, the University of Hawaii, the Pan-STARRS Project Office, the Max-Planck Society and its participating institutes, the Max Planck Institute for Astronomy, Heidelberg and the Max Planck Institute for Extraterrestrial Physics, Garching, The Johns Hopkins University, Durham University, the University of Edinburgh, Queen’s University Belfast, the Harvard-Smithsonian Center for Astrophysics, the Las Cumbres Observatory Global Telescope Network Incorporated, the National Central University of Taiwan, the Space Telescope Science Institute, the National Aeronautics and Space Administration under Grant No. NNX08AR22G issued through the Planetary Science Division of the NASA Science Mission Directorate, the National Science Foundation under Grant No. AST-1238877, the University of Maryland, and Eotvos Lorand University (ELTE) and the Los Alamos National Laboratory.

This research is based in part on data collected at the Subaru Telescope and retrieved from the HSC data archive system, which is operated by the Subaru Telescope and Astronomy Data Center at NAOJ.

\software{astropy \citep{robitaille2013astropy}, emcee \citep{foreman2013emcee}, lmfit \citep{newville2016lmfit}, scipy \citep{virtanen2020scipy}, SALT2 \citep{guy2007salt2}, SNID \citep{2007ApJ...666.1024B}, Matplotlib \citep{hunter2007matplotlib}.}

\begin{longrotatetable}
\begin{deluxetable*}{lllrrrrrrrll}
\tablecaption{Characteristics of Early-phase SNe Ia\label{tab:messier}}
\tablewidth{0pt}
\tabletypesize{\scriptsize}
\tablehead{\colhead{Name} & \colhead{Redshift} & \colhead{Distance Module} & \colhead{E$(B-V)_{MW}$} & \colhead{Host Extinction$^a$} & \colhead{$M_{B,\text{max}}$ Epoch} & \colhead{$M_{B,\text{max}}^b$} & \colhead{$\Delta m$15(B)} & \colhead{$\Delta{t_{m(-13)}}$$^c$} & \colhead{Filter$^d$} & \colhead{EEx} & \colhead{Data Source$^e$}}
\startdata
SN 2005cf & 0.006 & 32.51(0.33) & 0.09(0.010) & Y & 53533.50(0.30) & -19.39(0.33) & 1.12(0.03) & 17.35(0.05) & B & N & 1\\
SN 2009ig & 0.009 & 32.82(0.09) & 0.01(0.010) & Y & 55080.50(0.00) & -19.46(0.12) & 0.90(0.07) & 16.54(0.00) & B & N & 2\\
SN 2010jn & 0.025 & 35.20(0.03) & 0.39(0.000) & Y & 55495.80(0.10) & -17.95(0.04) & 0.90(0.00) & 16.60(0.50) & g & N & 3\\
SN 2011fe & 0.002 & 29.13(0.03) & 0.04(0.000) & Y & 55814.98(0.03) & -19.23(0.19) & 1.18(0.03) & 17.27(0.00) & r & N & 4\\
SN 2012cg & 0.002 & 30.90(0.30) & 0.20(0.050) & Y & 56081.30(0.50) & -19.62(0.08) & 0.86(0.02) & 17.60(0.24) & B & Y & 5\\
SN 2012fr & 0.005 & 31.27(0.05) & 0.03(0.000) & Y & 56243.00(0.30) & -19.49(0.06) & 0.85(0.00) & 16.57(0.16) & LSQgr & N & 6\\
SN 2013dy & 0.004 & 31.48(0.01) & 0.34(0.000) & Y & 56501.10(0.00) & -19.67(0.02) & 0.92(0.00) & 17.70(0.05) & clear & Y & 7\\
SN 2013gy & 0.014 & 33.68(0.09) & 0.15(0.060) & Y & 56648.50(0.10) & -19.32(0.16) & 1.23(0.06) & 17.22(0.04) & g & N & 8\\
SN 2015F & 0.027 & 31.89(0.04) & 0.21(0.030) & Y & 57106.45(0.09) & -19.43(0.11) & 1.26(0.10) & 17.50(0.18) & R & N & 9\\
SN 2017eck & 0.025 & 35.26(0.01) & 0.66(0.010) & N & 57899.49(0.09) & -18.90(0.02)$^*$ & 0.90(0.09) & 16.10(0.10) & o & N & 10\\
SN 2017cbv & 0.004 & 30.49(0.40) & 0.11(0.000) & Y & 57840.46(0.42) & -19.49(0.05) & 0.87(0.07) & 19.30(0.23) & U & Y & 11\\
SN 2017erp & 0.006 & 32.29(0.26) & 0.09(0.000) & Y & 57934.90(0.00) & -19.48(0.03) & 1.05(0.06) & 18.19(0.50) & g & Y & 12\\
SN 2017fgc & 0.008 & 32.51(0.11) & 0.20(0.070) & Y & 57959.40(0.40) & -19.32(0.13) & 1.05(0.07) & 17.28(0.01) & clear & N & 13\\
SN 2018agk & 0.026 & 35.43(0.00) & 0.14(0.050) & Y & 58204.17(0.00) & -18.95(0.00) & 1.07(0.00) & 17.34(0.14) & r & N & 14\\
SN 2018aoz & 0.006 & 31.75(0.15) & 0.09(0.000) & Y & 58221.41(0.00) & -19.32(0.00) & 1.11(0.01) & 14.45(0.20) & B & N & 15\\
SN 2018gv & 0.005 & 31.76(0.00) & 0.12(0.020) & Y & 58149.69(0.00) & -19.07(0.00) & 0.96(0.00) & 17.25(0.05) & B & N & 16\\
SN 2018kav & 0.033 & 35.76(0.05) & 0.04(0.000) & N & 58481.12(0.10) & -18.88(0.02)$^*$ & 0.98(0.12) &  &  & N & 10\\
SN 2018oh & 0.011 & 33.61(0.05) & 0.04(0.000) & Y & 58162.70(0.30) & -19.47(0.10) & 0.96(0.03) & 17.91(0.00) & g & Y & 17\\
SN 2018yu & 0.008 & 33.01(0.00) & 0.12(0.000) & Y & 58194.32(0.04) & -19.52(0.01) & 0.95(0.00) & 18.10(0.58) & g & Y & 18\\
SN 2019ein & 0.008 & 32.95(0.12) & 0.10(0.020) & Y & 58618.24(0.07) & -19.11(0.23) & 1.36(0.02) & 14.84(0.05) & B & N & 19\\
SN 2019np & 0.005 & 32.48(0.09) & 0.01(0.000) & Y & 58510.20(0.80) & -19.52(0.47) & 1.04(0.04) & 17.92(0.36) & r & Y & 20\\
SN 2020nlb & 0.020 & 31.26(0.05) & 0.06(0.000) & Y & 59042.10(0.04) & -19.11(0.20) & 1.28(0.02) & 17.55(0.30) & B & N & 21\\
SN 2021aefx & 0.005 & 31.16(0.23) & 0.00(0.000) & Y & 59546.67(0.00) & -19.63(0.02) & 0.90(0.02) & 18.34(0.50) & B & Y & 22\\
SN 2021hpr & 0.009 & 33.28(0.11) & 0.02(0.070) & Y & 59321.85(0.21) & -19.55(0.11) & 0.98(0.02) & 17.67(0.22) & B & Y & 23\\
SN 2021ujx & 0.020 & 34.69(0.20) & 0.19(0.000) & N & 59437.92(0.09) & -19.13(0.02)$^*$ & 1.42(0.09) & 16.07(0.03) & o & N & 10\\
SN 2022achv & 0.049 & 36.76(0.01) & 0.13(0.010) & N & 59927.24(0.11) & -19.11(0.02)$^*$ & 0.86(0.13) &  &  & N & 10\\
SN 2022aech & 0.03 & 35.58(0.00) & 0.00(0.000) & N & 59951.01(0.11) & -18.91(0.01)$^*$ & 0.92(0.17) & 16.91(0.72) &  & N & 10\\
SN 2022wpy & 0.015 & 34.16(0.01) & 0.11(0.010) & N & 59870.72(0.02) & -19.12(0.01)$^*$ & 1.10(0.02) & 16.46(0.10) & o & N & 10\\
SN 2023bzm & 0.050 & 36.80(0.01) & 0.08(0.010) & N & 60007.70(0.13) & -18.89(0.03)$^*$ & 0.98(0.15) &  &  & N & 10\\
SN 2023wrk & 0.010 & 33.42(0.15) & 0.03(0.020) & Y & 60269.45(0.1) & -19.51(0.15) & 0.87(0.02) & 17.60(0.11) & g & Y & 24\\
HSC17bmhk & 0.340 & 40.48(0.00) & 0.00(0.000) & N & 57799.69(0.40) & -19.14(0.03) & 1.04(0.01) &  &  & Y & 25\\
iPTF13dge & 0.016 & 34.50(0.07) & 0.03(0.040) & N & 56556.35(0.08) & -19.20(0.00) & 1.18(0.00) & 17.20(0.08) & r & N & 26\\
iPTF16abc & 0.023 & 35.00(0.40) & 0.00(0.000) & N & 57499.54(0.00) & -19.24(0.04) & 0.91(0.00) & 18.20(0.10) & g & Y & 27\\
LSQ12gpw & 0.058 & 37.11(0.01) & 0.07(0.001) & N & 56268.40(0.16) & -19.65(0.03) & 0.449(0.00) &  &  & N & 28\\
LSQ12hxx & 0.067 & 37.53(0.01) & 0.01(0.001) & N & 56289.80(0.09) & -19.36(0.03) & 0.94(0.13) &  &  & N & 28\\
LSQ13cpk & 0.032 & 35.83(0.01) & 0.02(0.001) & N & 56590.05(0.07) & -19.46(0.02) & 0.89(0.08) &  &  & N & 28\\
LSQ13des & 0.060 & 37.20(0.01) & 0.02(0.001) & N & 56638.90(0.08) & -18.99(0.03) & 0.82(0.09) &  &  & N & 28\\
LSQ13ry & 0.031 & 35.75(0.01) & 0.04(0.001) & N & 56394.90(0.06) & -19.31(0.02) & 1.09(0.06) &  &  & N & 28\\
PTF10accd & 0.031 & 35.73(0.01) & 0.12(0.001) & N & 55556.77(0.11) & -19.09(0.02) & 0.89(0.08) &  &  & N & 28\\
PTF10duz & 0.060 & 37.21(0.01) & 0.02(0.001) & N & 55285.23(0.16) & -19.23(0.02) & 1.19(0.23) &  &  & N & 28\\
PTF10hml & 0.054 & 36.98(0.01) & 0.01(0.001) & N & 55352.30(0.07) & -19.27(0.01) & 0.90(0.07) &  &  & N & 28\\
PTF10iyc & 0.055 & 37.02(0.01) & 0.01(0.001) & N & 55361.50(0.11) & -19.21(0.02) & 1.06(0.18) &  &  & N & 28\\
PTF11hub & 0.029 & 35.56(0.01) & 0.01(0.001) & N & 55770.15(0.11) & -18.88(0.01) & 1.60(0.21) & 15.97(0.01) & PTF48R & N & 28\\
PTF11qnr & 0.016 & 34.31(0.01) & 0.06(0.001) & N & 55902.30(0.38) & -19.02(0.03) & 1.65(0.56) & 16.50(1.00) & PTF48R & N & 28\\
PTF12gdq & 0.033 & 35.87(0.01) & 0.02(0.001) & N & 56116.40(0.10) & -18.89(0.01) & 1.28(0.16) &  &  & N & 28\\
MUSSES2022S & 0.170 & 39.26(0.00) & 0.18(0.020) & N & 59894.23(0.14) & -19.27(0.03)$^*$ & 1.38(0.17) &  &  & N & 29\\
MUSSES2022T & 0.210 & 40.07(0.00) & 0.17(0.030) & N & 59896.82(0.20) & -19.29(0.03)$^*$ & 1.19(0.19) &  &  & N & 29\\
MUSSES2111D & 0.132 & 38.95(0.00) & 0.19(0.010) & N & 59540.07(0.08) & -18.71(0.03)$^*$ & 1.25(0.09) &  &  & N & 29\\
ZTF18aayjvve & 0.048 & 36.60(0.10) & 0.017(0.000) & N & 58291.92(0.00) & -18.84(0.03) & 1.06(0.03) &  &  & Y & 30\\
ZTF18abdfazk & 0.084 & 37.90(0.40) & 0.01(0.000) & N & 58307.08(0.00) & -19.01(0.03) & 1.18(0.04) &  &  & Y & 30\\
ZTF18abwkczr & 0.027 & 35.38(0.00) & 0.64(0.010) & N & 58391.34(0.08) & -18.96(0.03)$^*$ & 1.28(0.08) & 16.99(0.26) & g & N & 31\\
ZTF19aaklqod & 0.038 & 35.95(0.13) & 0.09(0.000) & N & 58904.19(0.06) & -19.22(0.03)$^*$ & 1.54(0.11) &  &  & N & 31\\
ZTF19abpgggu & 0.050 & 36.71(0.00) & 0.26(0.010) & N & 58721.63(0.07) & -18.90(0.03)$^*$ & 1.09(0.09) &  &  & N & 31\\
ZTF19abuhlxk & 0.027 & 35.25(0.14) & 0.86(0.000) & N & 58740.69(0.04) & -18.71(0.03)$^*$ & 1.06(0.04) & 17.35(0.10) & g & N & 31\\
ZTF20abbhyxu & 0.031 & 35.52(0.13) & 0.03(0.000) & N & 59008.51(0.04) & -19.39(0.03)$^*$ & 1.18(0.05) &  &  & N & 31\\
ZTF20actddoh & 0.018 & 34.28(0.16) & 0.09(0.000) & N & 59193.41(0.04) & -19.27(0.03)$^*$ & 0.93(0.03) & 17.59(0.31) & g & N & 31\\
ZTF20acugxbm & 0.026 & 35.31(0.00) & 0.1(0.010) & N & 59199.24(0.05) & -19.10(0.03)$^*$ & 1.03(0.05) & 17.27(0.37) & r & N & 31\\
ZTF20acxkxxr & 0.016 & 34.26(0.00) & 0.20(0.010) & N & 59212.81(0.08) & -18.96(0.04)$^*$ & 1.16(0.13) & 17.17(0.10) & g & N & 31\\
ZTF20acxzqyp & 0.035 & 35.78(0.13) & 0.02(0.000) & N & 59217.64(0.06) & -19.75(0.03)$^*$ & 1.06(0.02) &  &  & N & 31\\
ZTF21aapehpx & 0.043 & 36.20(0.13) & 0.07(0.000) & N & 59309.60(0.08) & -19.09(0.03)$^*$ & 1.20(0.07) &  &  & N & 31\\
ZTF21aatlesr & 0.040 & 36.09(0.12) & 0.15(0.000) & N & 59327.61(0.06) & -19.71(0.03)$^*$ & 0.98(0.05) &  &  & N & 31\\
ZTF21abbliav & 0.056 & 36.81(0.12) & 0.26(0.000) & N & 59360.59(0.07) & -19.26(0.03)$^*$ & 1.48(0.07) &  &  & N & 31\\
ZTF21abdveqn & 0.018 & 34.31(0.14) & 0.58(0.000) & N & 59384.14(0.07) & -18.92(0.03)$^*$ & 1.05(0.08) & 15.50(0.10) & g & N & 31\\
ZTF21abgxudc & 0.027 & 35.33(0.00) & 0.12(0.010) & N & 59389.75(0.09) & -18.99(0.03)$^*$ & 1.57(0.13) & 15.04(0.57) & g & N & 31\\
ZTF22aawdqfp & 0.047 & 36.42(0.12) & 0.09(0.000) & N & 59801.03(0.05) & -19.15(0.03)$^*$ & 1.15(0.05) &  &  & N & 31\\
ZTF22abqkzuq & 0.030 & 35.47(0.14) & 0.10(0.000) & N & 59895.30(0.08) & -19.18(0.03)$^*$ & 0.94(0.07) &  &  & N & 31\\
ZTF23aahfxpr & 0.024 & 34.98(0.14) & 0.08(0.000) & N & 60073.94(0.06) & -19.02(0.03)$^*$ & 1.01(0.06) & 17.27(0.75) & g & N & 31\\
\enddata
\tablecomments{\\
$^a$ SNe with host-extinction corrections are marked by ``Y" by following the treatment from published papers. Host extinctions for remaining samples are ignored as the color effect has been considered for the M$_{B,max}$ derived by SALT2.\\
$^b$ M$_{B,max}$ derived by SALT2 are marked by star symbols.\\
$^c$ $\Delta{t_{m(-13)}}$ values of SNe Ia at $z<0.03$. \\
$^d$ Filters used for fitting $\Delta{t_{m(-13)}}$. \\
$^e$ (1) \citet{pastorello2007esc}, (2) \citet{Marion_2013}, (3) \citet{Hachinger_2013}, (4) \citet{Zhang_2016}, (5) \citet{Marion_2016}, (6) \citet{Zhang_2014},  (7) \citet{Pan_2015}, (8) \citet{Holmbo_2019}, (9) \citet{Cartier_2016}, (10) ATLAS, (11) \citet{Wee_2018}, (12) \citet{Brown_2019}, (13) \citet{Zeng_2021}, (14) \citet{Wang_2021}, (15) \citet{Ni_2022}, (16) \citet{Yang__2020}, (17) \citet{Li_2018}, (18) \citet{burke2022earlylightcurvestypeia}, (19) \citet{Kawabata_2020}, (20) \citet{Sai_2022}, (21) \citet{williams2024observationstypeiasupernova}, (22) \citet{ni2022infant}, (23) \citet{Lim_2023}, (24) \citet{liu2024earlytimeobservationssn2023wrk}, (25) \citet{Jiang_2020}, (26) \citet{Ferretti_2016}, (27) \citet{Miller_2018}, (28) \citet{firth2015rising}, (29) MUSSES, (30) \citet{Deckers_2022}, (31) ZTF.}
\end{deluxetable*}
\label{tab:events}
\end{longrotatetable}

\bibliography{sample631}{}
\bibliographystyle{aasjournal}

\clearpage

\end{document}